# Micro/Nano Motor Navigation and Localization via Deep Reinforcement Learning

Yuguang Yang,[a,b] Michael A. Bevan[b], Bo Li[a,*]

[a] *Institute of Biomechanics and Medical Engineering, Applied Mechanics Laboratory, Department of Engineering Mechanics, Tsinghua University, Beijing 100084, China*
[b] *Chemical & Biomolecular Engineering, Johns Hopkins University, Baltimore, MD 21218*

## Abstract

Efficient navigation and precise localization of Brownian micro/nano self-propelled motor particles within complex landscapes could enable future high-tech applications involving for example drug delivery, precision surgery, oil recovery, and environmental remediation. Here we employ a model-free deep reinforcement learning algorithm based on bio-inspired neural networks to enable different types of micro/nano motors to be continuously controlled to carry out complex navigation and localization tasks. Micro/nano motors with either tunable self-propelling speeds or orientations or both, are found to exhibit strikingly different dynamics. In particular, distinct control strategies are required to achieve effective navigation in free space and obstacle environments, as well as under time constraints. Our findings provide fundamental insights into active dynamics of Brownian particles controlled using artificial intelligence and could guide the design of motor and robot control systems with diverse application requirements.

[*] Corresponding author.
 E-mail address: libome@tsinghua.edu.cn (B. Li).

# 1. Introduction

In the past decade, there has been growing interest in engineering active particles for a diverse range of applications.[1-8] Active particles are designed to harvest energy to power translational motion and are envisioned as potential micro-/nano motors to carry out tasks in complex, hard-to-reach environments (e.g., mazes, blood vessels and porous media). The potential of such motors has been demonstrated in emerging applications like drug delivery, precision surgery, and environmental remediation.[1,3,9-15]

The ability of efficient navigation (move from one position to another) and precise localization (maintaining a position) of micro-/nano motors in complex environments plays a crucial role in deploying micro-/nano motors in applications.[16] Unlike macroscale motors, micro-/nano motors are often under actuated (i.e., not all degrees of freedom can be controlled), and common experimental realizations usually allows individual control on self-propulsion speed (via light, acoustics, etc.[17]) or propulsion direction (via magnetic fields[18]) or speeds and direction combined[19], but rarely both speed and direction independently. Additional hurdles to reliable control include Brownian motion that can significantly cause deviations from intended trajectories. Further considering the rich locomotion dynamics resulting from constituent materials (e.g., metal, polystyrene, etc.[17,20,21]), motor shapes (e.g., spheres[22], rods[23], and rationally tailored shapes[19]), and activation mechanisms (e.g., chemical catalysis[21] and external fields[24]), it is desirable to have a generic algorithm that addresses under-actuation and stochastic disturbances and is broadly suited for different motor designs and control objectives.

Strategies to realize efficient navigation and precise localization include empirical and approximate methods in relatively simple navigation scenarios[25,26] and a more formal algorithmic optimization framework we developed recently that could accommodate complex[27] and even unknown obstacle environments.[28] Particularly, in light of recent fast developments of artificial intelligence and deep learning technologies,[29-31] we recently addressed the navigation challenge in large-scale, unknown obstacle environments via a data-driven visual-based deep reinforcement learning (DRL) algorithm[28]. The DRL algorithm employs a bio-inspired neural network architecture that mimics the visual navigation system based on local neighborhood sensor information and equips the motor agent with intelligence to efficiently navigate unknown landscapes with random obstacle configurations. Despite its success for binary-activation self-propelled colloidal motors (on and off of self-propulsion), this DRL algorithm can only apply to motors with discrete control inputs, thus failing to meet the requirement of continuous control in applications like high precision localization.

In this work, we develop a flexible, generic DRL algorithm that allows continuous control of motors with different translational and rotational dynamics to carry out localization and navigation tasks. Leveraging this DRL algorithm, we investigate and compare navigation and localization strategies employed in different scenarios, which ultimately provides guidance on designing future autonomous micro-/nano motor systems. By varying the input information and reward signal structure, we demonstrate its capabilities to navigate in free space and obstacle environments and under additional arrival timing constraints. Our results shed light on DRL-controlled motor dynamics and also provide new route towards devising motor control systems able to cope with complicated and diverse tasks.

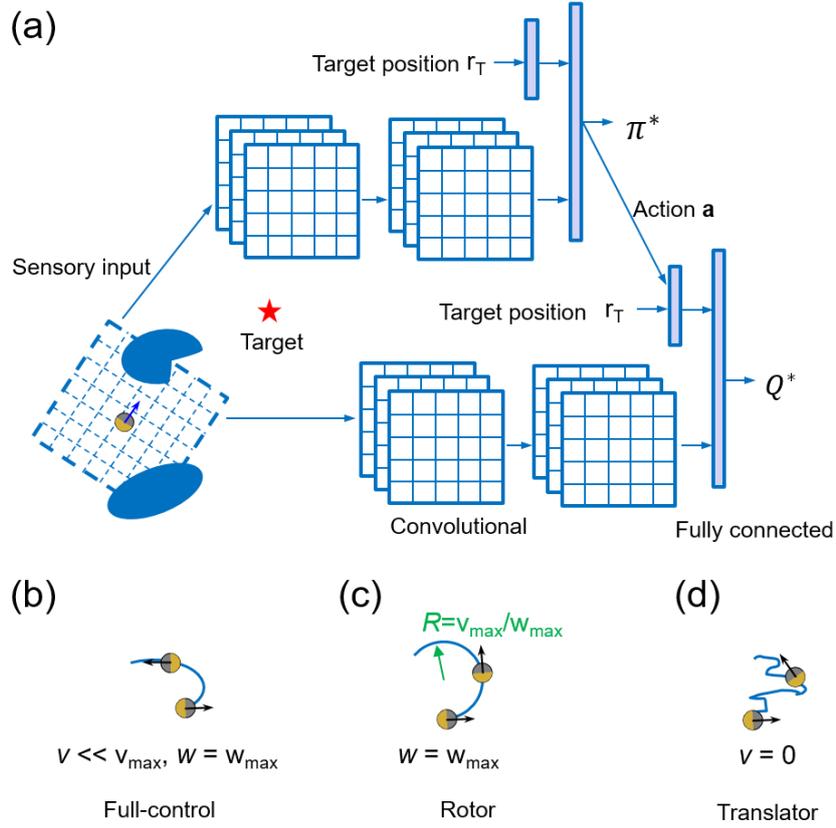

**Figure 1.** (a) The neural network architecture used in our DRL algorithm. The details of the architecture are provided in *Methods*. The neural network contains two sub-networks: an actor network and a critic network, where the actor network takes observation as input and outputs actions to adjust self-propulsion speeds and directions, and the critic network takes observation and actions as input and outputs corresponding Q value. The observation consists of two streams of sensory inputs, including pixel image ($30a \times 30a$) of the motor's neighborhood fed into convolutional layers and the target's position and actions fed into a fully connected layer. (b, c, d) Different types of motors we consider in this work and their re-orientation strategies. (b) A full-control motor with control of both its propulsion speed and direction will usually use a small self-propulsion speed and maximum rotation (i.e., $v \ll v_{max}$, $w = w_{max}$), v to achieve re-orientation. (c) A rotor motor with the sole control of direction will usually use maximum rotation (i.e., $w = w_{max}$) to achieve re-orientation. Because a rotor is constantly engaged in maximum self-propulsion ($v = v_{max}$), it will trace out a circular arc of radius $R = v_{max}/w_{max}$. (d) A translator motor with the sole control of propulsion speed with turn off the propulsion (i.e., $v = 0$) and wait for the desired direction sampled from Brownian rotation.

## 2. Models and Algorithms

In this work, we consider three types of motors, which have the basic locomotion elements among a wide range of motors. The first type of motor considered, which we refer to as a *full-control* motor hereafter, allows continuous control of its self-propulsion speed and direction. The second type of motor allows continuous control of its self-propulsion direction (e.g., via magnetic field[18]) but not its speed, which we refer to as a *rotor* motor. The third type of motor allows continuous control of its self-propulsion but not its orientation(e.g., via light[32]), which we refer to as a *translator* motor.

A full-control motor has the equation of motion given by

$$\partial_t \mathbf{r} = \xi_r(t) + \frac{D_t}{kT}\mathbf{F}\Delta t + v\mathbf{n}, \quad v \in [0, v_{\max}],$$
$$\partial_t \theta = \xi_\theta(t) + w, \quad w \in [-w_{\max}, w_{\max}], \tag{1}$$

a rotor motor has equation of motion of

$$\partial_t \mathbf{r} = \xi_r(t) + \frac{D_t}{kT}\mathbf{F}\Delta t + v_{\max}\mathbf{n},$$
$$\partial_t \theta = \xi_\theta(t) + w, \quad w \in [-w_{\max}, w_{\max}], \tag{2}$$

and a translator motor has the equation of motion of

$$\partial_t \mathbf{r} = \xi_r(t) + \frac{D_t}{kT}\cdot\mathbf{F}\Delta t + v\mathbf{n}, \quad v \in [0, v_{\max}]$$
$$\partial_t \theta = \xi_\theta(t), \tag{3}$$

where $\mathbf{r}=(x, y)$ and $\theta$ denote the position and orientation, respectively, $t$ is time, $kT$ is thermal energy, $\mathbf{F}$ is the force due to rod-obstacle electrostatic interactions (see *Methods*), and $v$ is propulsion speed taking binary values of 0 and $v_{\max}$ as the control inputs. Brownian translational and rotational displacement processes $\xi_r$ and $\xi_\theta$ are zero-mean Gaussian noise process with variances $\langle \xi_r(t)\xi_r(t')^T \rangle = 2D_t\delta(t-t')$ and $\langle \xi_\theta(t)\xi_\theta(t') \rangle = 2D_r\delta(t-t')$, respectively, where $D_t$ is the translational diffusivity and $D_r$ is the rotational diffusivity. All lengths are normalized by particle radius $a_R$ and time is normalized by characteristic Brownian rotational time $\tau = 1 / D_r$. The control update time is $t_c = 0.1\tau$, the integration time step $\Delta t = 0.001\tau$, $v_{\max} = 2\ a_R / t_c$, $D_t = 1.33a^2 D_r$, and $w_{\max}$ takes different values that will be specified in the following sections.

We formulate the tasks of localization and navigation as sequential decision-making processes in which a motor agent will be rewarded when it is sufficiently close to the specified target location. Formally, we use $s_n = (\mathbf{r}_n, \theta_n)$ to denote the motor's state, where the subscript $n$ is the indexed time step. The motor's observation at $s_n$, denoted by $\phi(s_n)$, is comprised of a binary image representation of the motor's square neighborhood and the target position ($\mathbf{r}^t$) in the motor's local frame, as shown in Fig. 1. We represent the motor's decision-making by control policy $\pi$, which maps an observation $\phi(s_n)$ to its decisions on self-propulsion and rotation, denoted by $a$. An optimal control policy $\pi^*$ that encourages the motor to localize itself around and navigate towards a specified target can be obtained by maximizing the expected reward accumulated during a navigation process, $\mathbb{E}\sum_{n=0}^{\infty}\gamma^n\left[r(s_{n+1})\right]$, where $r$ is the one-step reward function and $\gamma$ is the discount factor to reward rewards in future states. We set $\gamma = 0.99$ to encourage the learning of policies that value rewards coming from distant future. To minimize localization error and arrival time,[27,33] the reward $r$ is set equal to 1, where the motors locate within a threshold distance to the target and 0 otherwise (see *Methods* for additional details on setting up reward functions).

We use a deep neural network, known as an actor network, to approximate the optimal control policy and another deep neural network, called a critic network, to approximate the optimal state-action value function [Fig. 1], which is known as the $Q^*$ function. $Q^*$ function is given by

$$Q^*\left(\phi(s),v\right) = \mathbb{E}\left[r(s_1) + \gamma^1 r(s_2) + \gamma^2 r(s_3) + \cdots \mid \phi(s_0) = \phi(s), a_0 = a, \pi^*\right], \quad (4)$$

which is the expected sum of rewards along the process by following the optimal policy $\pi^*$, after observing $\phi(s)$ and an initial action $a$. Both neural networks employ convolution neural layers to process sensory information about the particle neighborhood, represented by a $W \times W$ binary image ($W = 30$), and a fully connected layer to process the target's position and actions.

We use a DRL algorithm known as deep deterministic gradient descent[34] plus additional enhancements[35,36] to simultaneously train the two networks to approximate their desired target functions. We train the neural network through extensive navigation data in different navigation scenarios with the goal to learn robust navigation strategies (see *Methods* for additional details on training neural networks).

## 3. Results and Discussions

### 3.1 Free space localization and navigation dynamics

We first examine the navigation and localization strategies obtained from our DRL algorithm for different types of motors in free space. Before we discuss the specific control policies for each type of motor, we first discuss the high-level mechanism that motors manage to get to targets located at different positions. For both navigation and localization, different motors control either propulsion speed or direction or both such that they can quickly move to specified targets. In targets are lying in front of them, the control strategies are relatively straight forward – simply self-propelling toward the targets. When targets are lying elsewhere, an adjustment on self-propulsion orientation are necessary. Fig. 1(b-d) schematically summarize the strategies employed by different types of motors to achieve major reorientation. A full-control motor will employ a small propulsion speed and maximum rotation to re-orient, analogous to steering an automobile [Fig. 1(b)]. A rotor motor will also engage the maximum rotation. Because a rotor is constantly engaged in maximum self-propulsion ($v = v_{max}$), it will trace out a circular arc of radius $R=v_{max}/w_{max}$ as it engages full rotation speed ($w = w_{max}$) for orientation adjustment [Fig. 1(c)]. A translator motor, due to its inability to directly control orientation, simply turns off self-propulsion and will wait for Brownian motion to sample the desired orientation [Fig. 1(d)]. The typical waiting time for a translator is thus on a scale of characteristic Brownian rotational time $\tau$ .

Fig. 2(a) and (b) show the control decisions for a full-control motor on propulsion and rotation speed (normalized by $v_{max}$ and $w_{max}$). The control decisions are parameterized by the different target locations while the motor is placed at the origin and orients along the *x* axis. Key aspects of the control strategy are summarized as following: (i) If the target exactly locates in front of the motor, ~zero rotation is applied and self-propulsion is employed, with the amount proportional to the distance up to $v_{max}$; (ii) If the target locates behind the motor, ~zero self-propulsion is applied but the maximum rotation speed (i.e, -1 and 1) is used to quickly reorient itself; (iii) When the target locates inside a wide vision cone with cone angle ~ ±120°, both rotation and propulsion are engaged, with the amount roughly in proportion to the distance and angle deviation; (iv) Even when the target is lying behind (vision cone angle 90~120°), nonzero propulsion is engaged to coordinate with the rotation to achieve the target as soon as possible.

The control strategies for a rotor motor [Fig. 2(c-e)] display similar structures to the rotation decision of the full-control motor but with additional structures depending on the ratio of $v_{max}$ over $w_{max}$. This ratio determines circular radius $R$ of the trajectory when a rotor motor employs $w_{max}$ for re-orientation [Fig. 1c]. When a target is locatedin the right front, orientation adjustment is unnecessary and thus zero rotation is applied; when a target locates in the right back, maximum rotation is applied for prompt re-orientation. When the target is lying front but with some angle off, the rotor applies rotation, increasing with the angel deviation, to re-orient itself, but has one critical difference to full-control motor: The rotor usually applies larger rotations in order to quickly re-orient itself since the maximum propulsion speed is always engaged, whereas for full-control motor, its rotation and propulsion are well coordinated to orient and move toward the target.

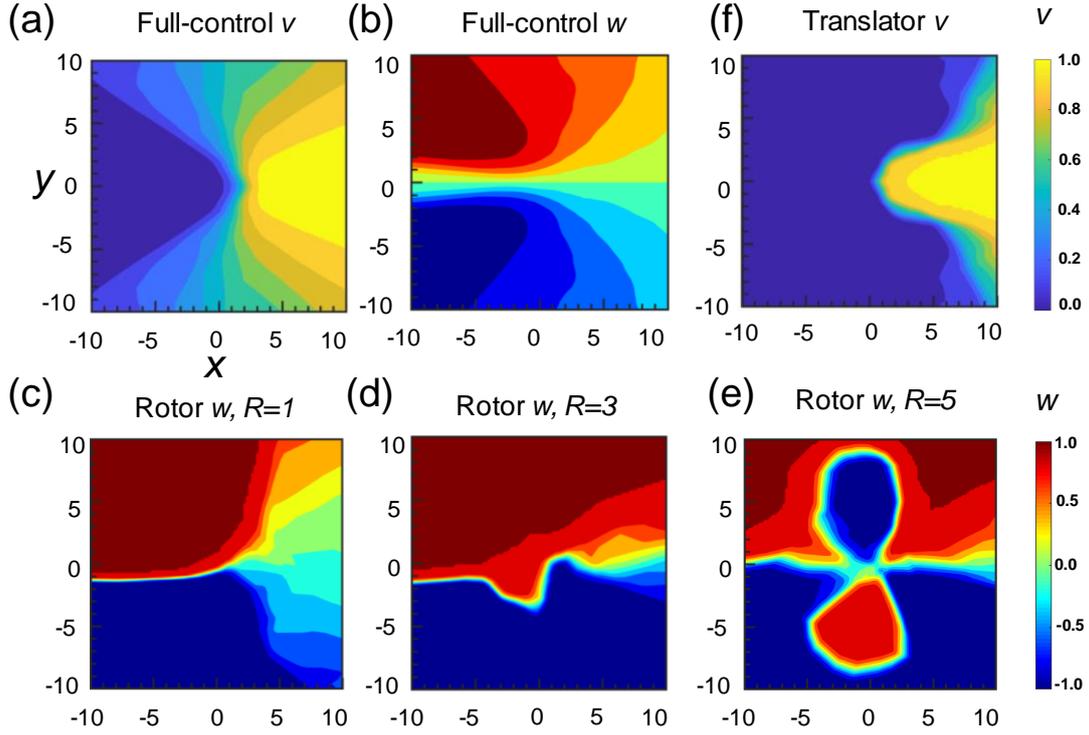

**Figure 2.** Learned control strategies for different types motors and representative controlled trajectories in free space navigation. In presenting the control policies, we place the motor at (0, 0) with orientation aligning with $x$ axis and vary the target location. (a, b) Normalized control strategies (normalized by $v_{max}$ and $w_{max}$) of propulsion speed (a) and rotation speed (b) as a function of target locations for a full-control motor. (c, d, e) Normalized control strategy of rotation speed for rotor motors with circular radius $R=1$, $R=3$, and $R=5$, respectively, with $R= v_{max}/w_{max}$ (e) Normalized control strategy of propulsion speed for a translator motor.

Additional structures emerge the control policies [Fig. 2(c-e)] when the target locates near the two sides of the motor with a large $R$. Because a rotor motor is constantly engaging the maximum self-propulsion, it cannot directly arrive at target on their near side by simply changing orientations to the side where the target lies. Instead, the rotor will first re-orient to the other direction to temporarily move away from the target, which can be rationalized by the need to gain more room to re-orient. As we increase allowable maximum rotation speed $w_{max}$ (i.e., decreases the circular radius), the control strategy converges to the full-control case.

The optimal control policy of translator motors can be coarsely summarized as orientation timing; that is, self-propulsion is on when the motor favorably orients to target and off if their orientation is unfavorable. The strength of self-propulsion is approximately proportional to the distance between the target and the motor, up to $v_{max}$. Similar strategies have been revealed in a number of previous studies[25,27,28,37,38].

Navigation trajectories of different motors under control steered towards targets at different locations are shown in Fig. 3(a-e). Full-control motors employ a combination of propulsion and rotation strategies, as shown in Fig. 2(a) and (b), to realize efficient navigation towards and localization around specified targets [Fig. 3(a)]. Without Brownian motion, trajectories are initialized with re-orientation towards the target if needed and continue with subsequent straight-line movement; with Brownian motion, the rotation will be constantly employed to correct orientation deviations and leads to curved trajectories. After arrival, full-control motors localize themselves by simply applying ~zero propulsion and rotation in absence of Brownian motion or employ the same strategies in Fig. 2(a) and (b) to correct deviations from Brownian motion.

The navigation and localization trajectories of rotor motors [Fig. 3(b-d)] display several interesting features compared to full-control motors. When a rotor motor navigates to targets in the back or on the side, its trajectory will trace out arcs with larger radius (i.e., it needs more room to re-orient) compared to full-control motors (see upper panel of Fig. 3(a) and (b-d)) due to their constant-on maximum self-propulsion. Rotor motors also display interesting localization behaviors as a result of inability to control its propulsion. Because the propulsion is constantly engaged, after passing through the target, the rotor still needs to constantly adjust its orientation in order to get back to its target. As a result, their trajectories can form regular patterns surrounding the target in the zero-noise limit or irregular ones when there is Brownian motion.

Compared to full-control and rotor motors that can directly control orientation by rotation, translator motors rely on Brownian motion to sample favorable directions. Controlled trajectories of translator motors demonstrate an intermittent, non-smooth features since they need to stop and wait for the favorable orientation from Brownian rotation [Fig. 3(e)].

We further compare the navigation and localization dynamics of different motors by examining their distance versus time as they navigate towards and localize around a target in front of them and at a distance of $10a$ [Fig. 3(f) and (g)]. Full-control and rotor motors can first arrive at the target around $0.5\tau$ (the minimum time possible) as they directly head towards the target at the maximum speed. The translator motors first quickly propel $\sim 5a$ towards the target as their initial orientations are favorably oriented. Then they stay around with no propulsion and waiting for favorable orientation to be sampled by Brownian rotation and finally arrive at the target at around $4\tau$. After arrival, full-control motors can closely localize around the target, with the motor-target distance vanishing in absence of Brownian motion and $\sim 1a$ in presence of Brownian motion. Rotor motors will periodically circulate around the target, with the maximum distance $\sim 2R$. Although translator motors arrive at target substantially slower than rotor motors, they can stay around the target with a distance of $\sim 1a$ by turning down propulsion strength.

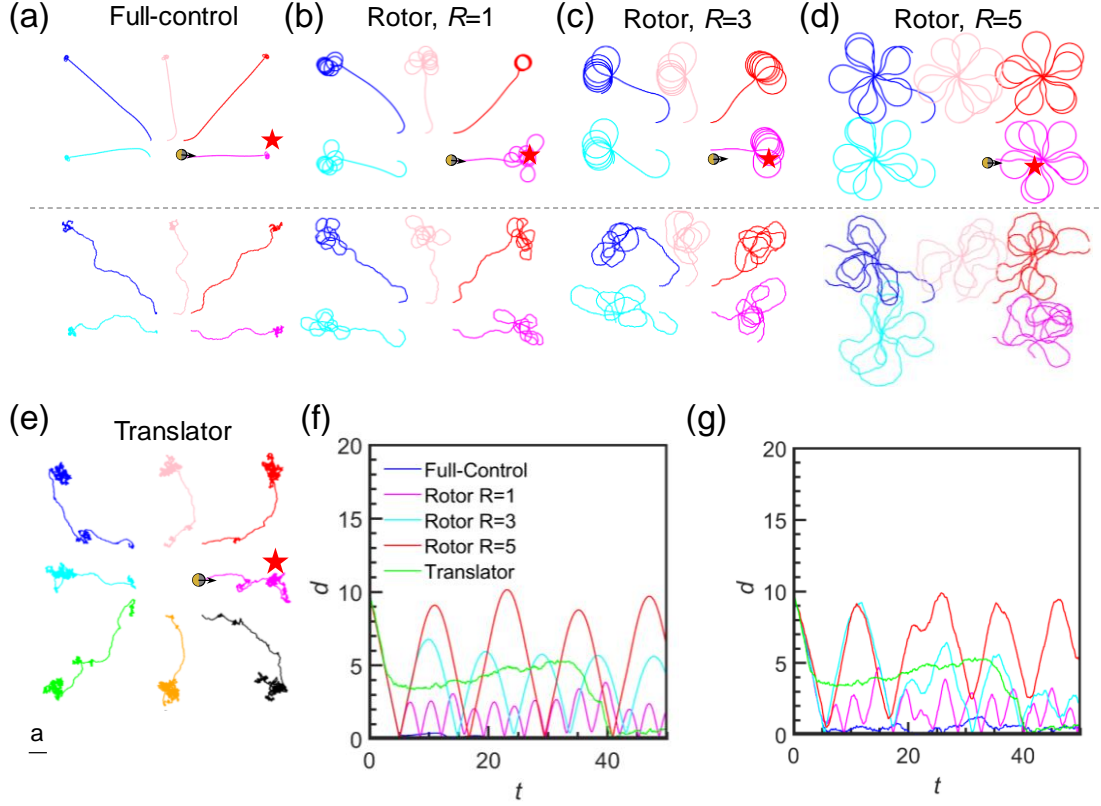

**Figure 3.** (a-e) Navigation and localization trajectories of motors staring at different location but with the same horizontal orientation and towards different targets, denoted by stars (see Methods for more details on the setup). In (a-d), upper panels (above the dashed line) are trajectories without Brownian translation and rotation for the purpose of illustrating the control policy; lower panels (below the dashed line) are trajectories with Brownian translation and rotation. (f, g) Motor-target distance versus time as motors navigate towards and then localize around a target in front of them and at a distance of $10a$. To more clearly illustrate the navigation and localization dynamics, Brownian translation and rotation are not added to full-control and rotor motors in (f), while they are added in (g).

## 3.2 Free space localization and navigation performance

We now quantify the navigation performance by comparing the mean traveled distance of motors within given time when they are controlled to transport along a fixed direction. Fig. 4(a-b) shows their traveled distance versus time within a fixed period of $50\tau$ as they are navigating along the horizon direction. Representative trajectories in Fig. 4(a) show that Brownian motion deviates motors' navigation trajectories towards their horizontal remote target, their propulsion and rotation decision largely maintain themselves near the ideal horizontal transport path. Full-control motors have transport speed ~$0.85v_{max}$, 15% lower than ideal navigation speed $v_{max}$ owing to Brownian motion disturbance. Rotor motors are slightly slower than full-control motors, particularly for $R = 5$ rotor motors due to small rotation capability unable to correct deviations from Brownian motion that slows down navigation.

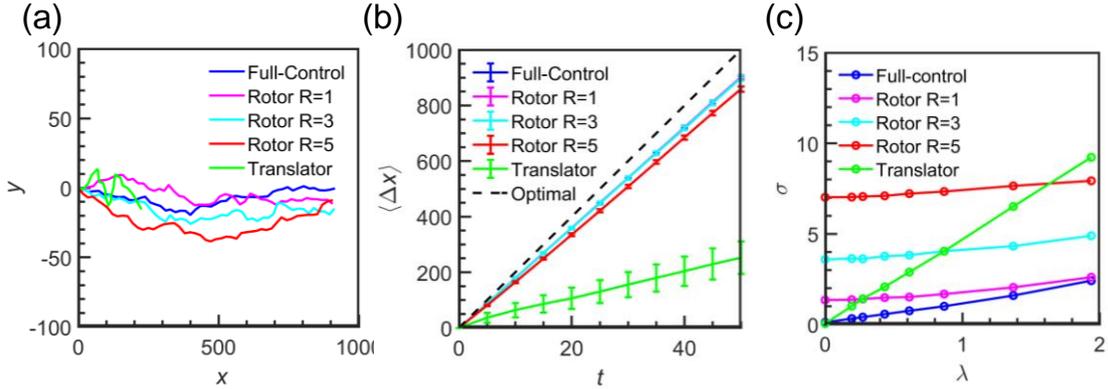

**Figure 4.** (a) Navigation trajectories lasting $50\tau$ of different motors starting initial state (0, 0, 0) toward a target located at (1000, 0). (b) Navigation performance of different types of motors characterized by mean travelled distance versus time along the horizontal distance. Dashed lines are optimal navigation at speed $v_{max}$. (c) Location performance of different types of motors characterized by the steady state deviation $\sigma$ from target as a function of position perturbation strength $\lambda$.

Translator motors have the worst navigation speed ~$0.23v_{max}$, which agrees with a theoretical approximate $v_{max}\int_{-\theta_c}^{\theta_c}\cos(\theta)p^{eq}(\theta)d\theta \approx 0.225v_{max}$ where $\theta$ is the orientational angle, $p^{eq}(\theta)= 1/2\pi$ is the equilibrium distribution of orientational angle, $\theta_c = \pi/4$ is estimated from control policy in Fig. 4(b). Moreover, translator rotors have a substantially larger standard deviation in its travelled distance (see error bars in Fig. 4(b)), indicating its lack of reliability to arrive in time compared to the other two types of motors. The substantial navigation inefficiency in translator motor is attributed to its reliance on Brownian motion to adjust its orientation to favorable regions. Results in Fig. 4(b) demonstrate that the controllability on self-propulsion direction plays a much more critical role in long-distance navigation than the controllability on self-propulsion speeds.

We further examine the localization performance of motors under various strengths of external disturbance imposed on motors' positions. The localization performance is characterized by the steady-state motor-target distance

$$\sigma = \langle \|\mathbf{r} - \mathbf{r}_\mathrm{T}\| \rangle, \tag{5}$$

where the bracket indicates evaluation using samples drawn from steady state (see *Methods*). Increasing the strength of external disturbance on motors' positions is realized by increasing the translation diffusivity $D_\mathrm{t}$ in Eqs. (1)-(3). We characterize the disturbance strength by a non-dimensional parameter $\lambda$

$$\lambda = \frac{\sqrt{D_\mathrm{t}\Delta t_C}}{v_\mathrm{max}\Delta t_C}, \tag{6}$$

where $\lambda$ is the ratio of random displacement over self-propulsion distance within one control time step $\Delta t_C$. Notably, in estimating $\sigma$ at various levels of $\lambda$, we only increase $\lambda$ to ~1 as further increasing position disturbance will simply lead to predominantly random walk and the steady state will be unattainable.

As shown in Fig. 4(c), full-control motors display the best localization performance over the whole range of $\lambda$. Although rotor motors have similar navigation performance to full-control motors, their localization performance is the worst among all types of motors at small $\lambda$, particularly for rotor motors with large circular radius $R$. Because of non-controllability on self-propulsion, rotor motors rely on the hovering strategies for localization. Hovering with large circular radio $R$ can cause proportional large deviation to the target [Fig. 3(f) and (g)]. Finally, a translator motor has an intermediate localization performance at small $\lambda$, thanks to its ability to turn off propulsion when not needed.

As we increase $\lambda$, localization errors for all types of motors increase, but at different speeds. Particularly, localization errors of translator motors increase linearly and at a much faster speed compared to that of full-control motors. Localization errors of rotor motors increase at relatively slow speeds because they initially have relatively large errors already. At larger $\lambda\sim1$, the rotor motor starts to outperform translator motor and the performance gap between full-control motor and rotor motor narrows. This is because as the random displacement at one-control step is comparable to the propulsion distance, the localization problem reduces to a free space navigation problem, and thereby the importance of direction control outweighs the propulsion control, as we concluded from Fig. 4(b).

In short, results in Fig. 4(c) demonstrate that: (i) At $\lambda \ll 1$, the localization performance is primarily impacted by the controllability on self-propulsion speed; (ii) At larger $\lambda$, the localization performance is impacted by the controllability on self-propulsion speed and direction, with the latter playing an increasingly predominant role.

### 3.3 Obstacle environment navigation

After understanding the navigation and localization in the free space, we now consider navigation strategies of different types of motors in environments with obstacles. We consider long narrow channel environments where obstacles are placed in the middle and on the side to block the motors [Fig. 5(a-d)]. Efficient navigation in the channel requires the motor to circumvent the obstacles in the middle lane and quickly get out of concave traps formed by the obstacles on the side and the walls. The obstacle channels are spacious enough for rotor motor $R$=1 to gracefully turn around but not enough for rotor motor $R$=5. We consider the navigation environments with both convex squares and concave crosses obstacles to perform finer examination of navigation capability under different circumstances.

Representative controlled navigation trajectories for different motors inside obstacle environments are shown in Fig. 5(a-d). In both square and cross obstacle channels, full-control motors [Fig. 4(a)] can swiftly control their self-propulsion direction to successfully get around obstacles and avoid getting trapped by concave geometries. The resulting trajectories usually closely follow the boundaries of the obstacles in the middle of the channels to shorten travel path distance for faster arrival. Rotor motors with a small circular radius $R$=1 [Fig. 4(b)] display similar navigation behavior to full-control motors since they both have the capability of fast adjusting orientation to avoid obstacles and traps. On the other hand, rotor motors with a large circular radius $R$=5 [Fig. 4(c)] can only adjust direction slowly, and thereby their navigation process usually involves accidentally hitting on the obstacle wall while adjusting the directions. Another downside of the slow direction adjustment is the resulting larger trajectory excursions away from the boundaries of middle lane obstacles, causing delayed arrivals. In addition, cross obstacles can often temporarily trap rotor motors with $R$=5 since they cannot move away from traps in an agile manner.

Because translator motors have no direct control on propulsion direction, they have to wait (at ~zero propulsion) for desired orientations sampled from Brownian rotation and then self-propel to circumvent obstacles when favorable directions are sampled. The trajectories of a translator motor [Fig. 4(d)] have similar features to that of a rotor motor with $R$=5, namely, large deviations from middle lane obstacle boundaries and occasional trapping by cross obstacles.

We perform more quantitative comparison [Fig. 5(f)] on the navigation performance by comparing the horizontal distance travelled versus time in infinitely extended patterned channels like Fig. 5(a). For all motors, the mean distances traveled versus time is linear, with an average speed ~$0.65v_{max}$, dropping by ~25% from ~$0.85v_{max}$ in free space navigation. The linearity in travelled distance versus time, instead of leveling off, indicates that controlled motors can all successfully pass through obstacle channels.

Full-control motors and rotor motors with $R = 1$ and 3 have similar performances, irrespective of obstacle shapes. Rotor motors with large circular radius $R$=5 cannot make prompt turns, causing them to get into traps and travel shorter distance (in terms of horizontal distance) within given time. Translator motors display an average navigation speed of ~$0.13v_{max}$, a drop of ~40% with respect to its free space navigation performance. Compared with the relatively small drop of ~25% of full-control and rotor motors in presence of obstacles, translator motors' navigation performance is more markedly affected by the presence of obstacles. Another noticeable result in Fig. 5(e) is that the standard deviations of traveled distance for translator motors are much smaller than that in free space navigation due to the confinement effects of obstacles.

We also find that cross obstacles can lead to slower navigation speed for all types of motors. In general, increased proportions of concave features will tend to trap all types of motors, but it only marginally impacts full-control motors and rotor motors with $R$=1. More remarkable impacts from concave geometry are found for rotor motors with $R$=5 and translator motors, where the former cannot turn around quickly due to rotation speed limit and the latter cannot directly control direction. Translator motors are affected the most because concave cross obstacles require re-orientation to a larger extend than square obstacles do, and thus require more waiting time for Brownian rotation sampling. We further characterize the effect of concave geometry on navigation of translator motors via first passage time distribution comparison in Fig. 5(f). Clearly, increasing convex features of obstacles not only can increase the mean first passage time, but also lead to substantial heavier tails in the distribution resulting from the trapping effects.

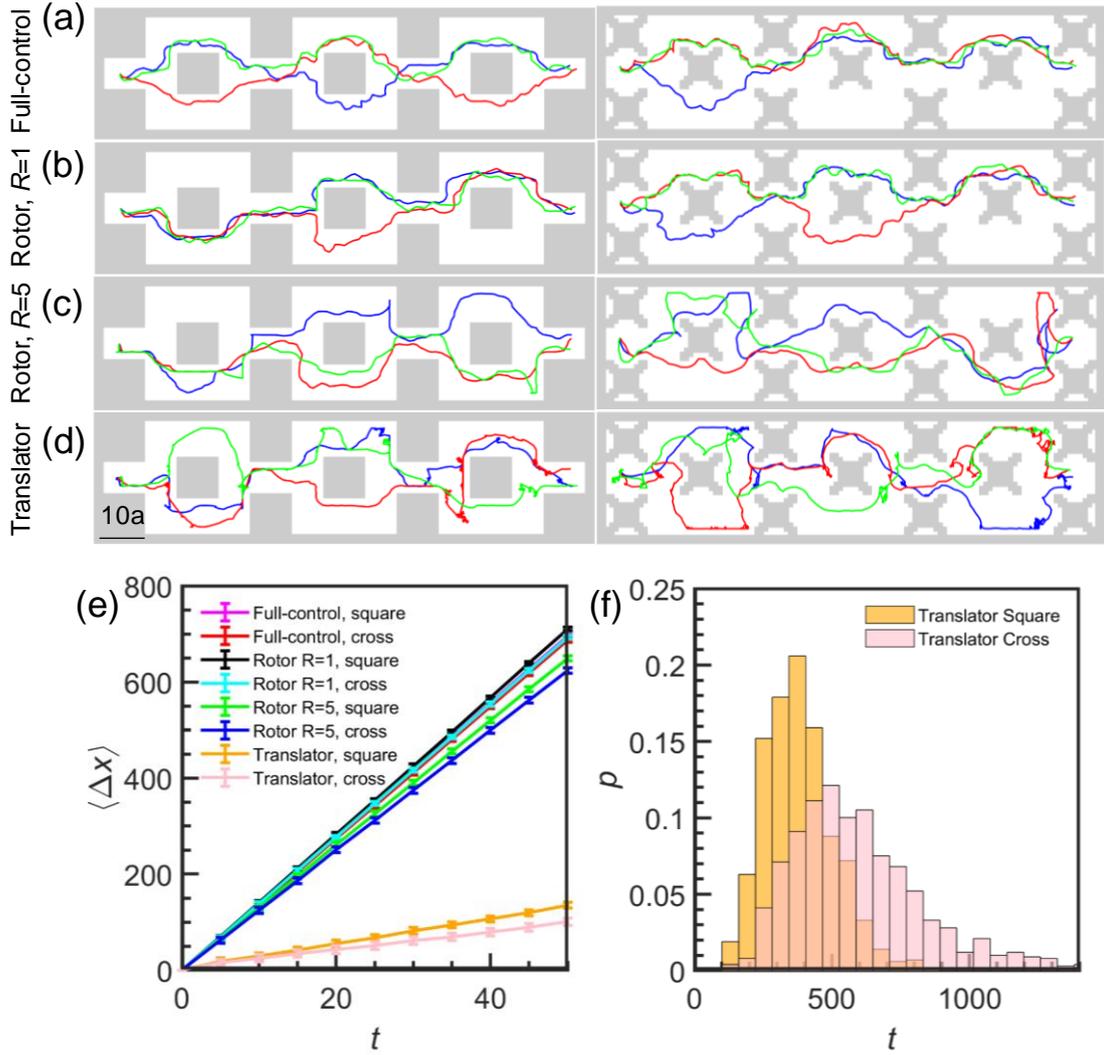

**Figure 5.** (a-d) Controlled trajectories of different motors navigating through channels filled with square (left) and cross obstacles (right). Motors are starting at the left end and navigating towards the right end of the channel (e) Navigation performance in the obstacle channel of different types of motors characterized by the mean travelled distance versus time along the horizontal distance. (f) First passage time distribution of a translator motor in the square obstacle channel versus the cross obstacle channel.

## 3.4 Temporal control

In previous examples, we have investigated the control of motors in space, where motors are navigating towards or localizing around specified spatial targets. The flexibility of our DRL algorithm also allows other control objectives, such as control in the temporal dimension, with minimal modifications on the input and reward function in the algorithm. Here we consider an example of arrival time control objective where we require the motors to arrive at specified locations within a specified time window, neither sooner nor later. Such arrival time control capability could be of potential use for motor applications with timing constraints. For example, in automatically scheduled drug release, drugs are required to be delivered within a restricted time window. More broadly, additional temporal control could enable solutions to problems involving collective dynamics where individuals are precisely controlled to synchronise and coordinate in time (e.g., ants, colloidal swarms).[39-41]

We achieve arrival time control by including time as an observation variable (together with the target location) and provide a time-dependent reward signal that encourages arrivals within the time window but discourages arrivals at other times. For demonstration purpose, we consider motors that navigate to a specified location in free space but requires earliest arrival after $T_c = 5\tau$. We set reward of $r = 1$ for arrivals after $T_c$ and $r = -1$ if arrivals are earlier than $T_c$, aiming to penalize early arrivals. Because the motor receives discounted rewards (i.e., via $\gamma < 1$), the control policy will be optimized to steer motors to specified target as early as possible after $T_c$.

Fig. 6(a-c) show representative navigation trajectories of different motors from the origin (0, 0) to a target at (20, 20), with arrival time constraints applied. We select such short-ranged target (distance $\sim 28a$) that motors can mostly arrive earlier than the allowed time $T_c$. Different motors have learned different navigation strategies that accommodates an arrival time constraint. The full-control motor employs a slow-down strategy that it reduces its self-propulsion speed and slowly arrives at the target at the required time windows [Fig. 6(a)]. The rotor motor cannot control its self-propulsion speed; instead, it will first steer towards the vicinity of the target and then hovers around the target as part of postponing its arrival until $T_C$ [Fig. 6(b)]. Translator motors will first engage their full power to get to the vicinity of the target and then wait till $T_c$, after which they self-propel right away to the target [Fig. 6(c)]. On a higher level, the rotor and translator motors are taking a similar early-arrive-and-wait strategy but implement it according to their specific dynamics. Notably, navigation strategies that satisfies the arrival time constraint are not unique. For example, instead of taking the slow-down strategy, the full-control motor can also first wait somewhere and then employ full

power. Our algorithm usually tends to find local optimal solutions that give relatively smooth strategies (in terms of variations of self-propulsion speeds and directions).

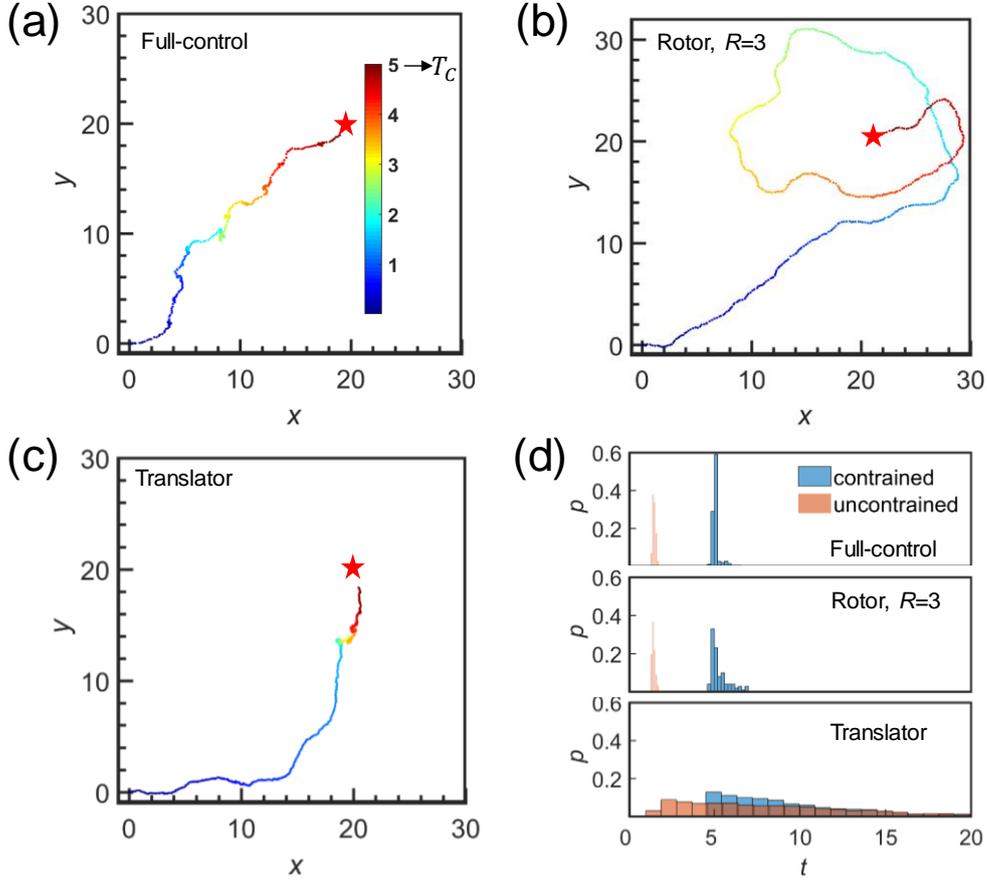

**Figure 6.** Representative navigation trajectories of motors with arrival time constraint for a full-control motor (a), rotor motor with $R = 5$ (b), and a translator motor (c). Motors are controlled to arrive at the earliest time after $T_C = 5\tau$. (d) The histogram of arrival times of controlled navigations with and without arrival time constraint (motors are controlled to arrive as early as possible).

To understand the underlying rationale for these adapted decisions, we further quantify the first arrival time statistics [Fig. 6(d)] for motors with constrained versus unconstrained arrival times. Without arrival time constraints, the full-control and the rotor motors arrive around 1.5 $\tau$; with arrival time constraints, full-control and rotor motors arrive around the allowed time $T_c$. Without arrival time control, the translator motor has a wide heavy-tailed arrival time distribution, with its mode around $3\tau$. The wide distribution arises from sampling of favourable orientations via Brownian rotation. Statistically, they have a significant chance of arriving very late if not enough favourable orientations are sampled. After adding the constraint, the translator motor's arrival time has a sharp peak at $T_c$, with a similar tail to the unconstrained case. The formation of peaks for translator motors is the result of the early-arrival-and-wait strategy where large portion of motors take immediate action to arrive at the target near $T_c$. Notably, the addition of arrival time constraint does not cause a heavy tail after >10τ, indicating that the constraints only push back early arrivals but do not affect late arrivals in the original unconstrained setting. An interesting aspect on the strategy of translator motor is that the translator motor does not adopt the simple slow-down strategy like the full-control motor. This is because such a slow-down strategy will push back all trajectories and is suboptimal as it delays late arrivals further.

In short, adding arrival time control objectives regulates the learned strategies. This control strategy is the compromised result of the arrival time requirement and the inherent uncontrollable elements of the motor dynamics. In terms of application guidelines, the full control motor and rotator can achieve a good arrival time control, although the rotor will require additional hovering space, which will probably become an issue for applications involving strong confinement.

## 4. Conclusions and outlook

We developed a general DRL algorithm that enables continuous control of a broad class of micro/nano motors in a number of navigation scenarios including free space, obstacle environments, and arrival time constraints. Our DRL can learn competitive strategies solely through navigation data without knowledge of the underlying model. Different motor locomotion dynamics and control objectives have led to different control strategies in free space and obstacle environments. Although it seems a lack of control degree of freedom might significantly impair its functions and performance, our DRL is able to alleviate negative impacts by employing different control strategies in navigation and localization in free space, obstacle environment, and navigation timing control.

Our DRL algorithm is model-free in that can be used to realize continuous control on micro/nano motors with other dynamic models[1,22,42-45] (e.g., other actuation mechanisms, hydrodynamics, etc.) beyond what has been considered here. Further extension of our DRL includes feeding other visual cues in the neural network such that motors can learn navigation and localization in environments like flow fields[19] and 3D porous media. Our DRL algorithm can also be conveniently integrated with experimental systems because of its capability to directly process raw sensor inputs from experimental imaging systems (e.g., microscopes).[46] More broadly, our algorithm can also serve as a general-purpose algorithm for diverse continuous control tasks on the microscopic scale, such as controlling colloidal assembly on energy landscapes,[47-50] that challenges the classical Markov decision process controller[51] because of the curse of dimensionality.

Various comparison studies in this work also provide useful directions to the design of navigation and motor systems for micro/nano robots. The full-control motors have the best performance in all the navigation and localization tests, suggesting that the hardware design of two control degrees of freedom is a direction worth pursuing, despite its considerable challenge. The continuously controllable rotor motors and translator motors are the most accessible experimental designs since only one control degree of freedom is needed. Particularly, if we allow enough maximum rotation speeds and sufficiently fast control frequency, a rotor motor can function comparable to a full-control motor. To realize both advantages of translator and rotor motors with only one control degree of freedom, one can use motors[37] with the switching ability between translator motors and rotor motors. A translator motor is considerably disadvantageous in long-distance navigation and obstacle environment navigation, although they have reasonable localization performance. A potential route for improving the performance of translator motors will be exploiting the swarm intelligence, which can be achieved using multi-agent stochastic feedback control.[41]

## 5. Methods

## 5.1 DRL algorithm and training

*5.1.1 Obstacle representation and collision dynamics*

We directly convert environment maps to pixel images (pixel size 1$a$) using image processing software. Obstacle regions have value 1 whereas free space regions have value 0. The local neighborhood sensory input is obtained by first constructing a squared window of width $W=30a$ centering on the motor and aligned with its orientation and then extracting a 30 by 30 binary matrix from the environment maps. Same to our previous work,[28] we project distant targets (target with distance larger than 30$a$) to a proxy one located on a circle of radius 30$a$ centering on the motor. Target positions are represented in local coordinate system of the motor.

Obstacles on each pixel are represented by repulsive spheres to capture the interaction between the motor and the obstacles, whose interaction force [used in Eqs. (1)–(3)] are modeled by electrostatic repulsion, given as,[43,52]

$$\mathbf{F} = \frac{\mathbf{r}_{RO}}{r_{RO}} B^{pp} \kappa \exp\left[-\kappa(r_{RO} - 2a)\right], \quad (7)$$

where $\mathbf{F}$ is the force on the motor, $\mathbf{r}_{RO}=\mathbf{r}_O - \mathbf{r}$ with $\mathbf{r}_O$ being the position of the obstacle, and $r_{RO} = \|\mathbf{r}_{RO}\|$. $B^{pp}$ is the pre-factor for electrostatic interactions and $\kappa^{-1}$ is the Debye length. We use $B^{pp} = 2.2974a/kT$ and $\kappa^{-1} = 30$nm.

*5.1.2 Actor network*

The neighborhood sensory input first enters a convolutional layer consisting of 32 filters with kernel size 2×2, stride 1, and padding of 1, following a batch normalization layer, a rectifier nonlinearity and a 2×2 of maximum pooling layer. The output then enters a second convolutional layer consisting of 64 filters and the same kernel, stride and padding as the previous layer, following similarly by a batch normalization layer, a rectifier nonlinearly and a maximum pooling layer. The local target coordinate first enters a fully connected layer consisting of 32 units following by rectifier nonlinearity. Then the output from the target coordinate input and the sensory input will merge and enter a fully connected layer of 64 unit followed by rectifier nonlinearity. The output layer is a fully-connected linear layer with two output of normalized *w'* and *v'*. Note that tanh nonlinearity is applied to the output constrain the *w'* between [-1, 1] and sigmoid nonlinearity is applied to constrain *v'* between [0, 1]. *w'* and *v'* are then multiplied by $v_{max}$ and $w_{max}$ to get the final action output.

*5.1.3 Critic network*

Besides the target and neighborhood sensory input, action outputs from actor network will also be fed into the critic network. The neighborhood sensory input will pass through the same convolutional layers as in the actor network. The target input will first concatenate with the action output from the actor network. The concatenated vector then will enter a fully connected layer consisting of 32 units followed by rectifier nonlinearity. Then the output from the target coordinate input and the sensory input will merge and enter a fully connected layer of 64 unit following by rectifier nonlinearity. The output layer is a fully-connected linear layer with one output as the *Q* value given input of observation and action.

*5.1.4 Training algorithm*

The algorithm we used to the train the agent is the deep deterministic policy gradient algorithm[34] plus the hindsight experience replay enhancements[28,36] and scheduled multi-stage learning following the idea of curriculum learning.[53] At the beginning of each episode, the initial motor state and the target position are randomly generated in such a way that their distance gradually increases from a small value. More formally, let *D(k)* denote the maximum distance between the generated initial state position and target position at training episode *k*, which is given by

$$D(k) = S_m \times (T_e + (T_e - T_s)\exp(-k/T_d)), \qquad (8)$$

where $S_m$ is the maximum of width and height of the training environment (at free space we set $S_m = 100a$), $T_s$ is initial threshold, $T_e$ is the final threshold, and $T_d$ is the threshold decay parameter. Then during the training process, the motor gradually acquires control strategies of increasing difficulties (in terms of initial distance to the target).

During the training process, we add noises to the actions from actor network to enhance the exploration in the policy space. The noise is sampled from an OU process (on each dimension) given by

$$d\eta = -\alpha(m-\eta)dt + \sigma_{OU} dB_t \qquad (9)$$

where α is the reversion parameter, $m$ is the mean level parameter, $\sigma_{OU}$ is the volatility parameter. and $B_t$ is the standard Brownian motion process.

The complete algorithm is given below.

---

**Algorithm:** deep deterministic policy gradient **with hindsight experience replay**

Initialize replay memory $M$ to capacity $N_M$
Initialize actor network μ with random weight $\theta^\mu$ and critic network $Q$ with random weights $\theta^Q$
Initialize target actor network μ' and critic network $Q'$ with random weights $\theta^{\mu'}$ and $\theta^{Q'}$
**For** episode 1, $N_E$ **do**
  Initialize particle state $s_0$ and target position
  Obtain initial observation $\phi(s_1)$
  **For** n =1, maxStep **do**
    Select an action $a_n$ from actor network plus additional perturbation sample from an OU process.
    Execute action $a_n$ using MCMC simulation and observe new state $s_{n+1}$ and reward $r(s_{n+1})$
    Generate observation state $\phi(s_{n+1})$ at state $s_{n+1}$
    Store transition $(\phi(s_n), a_n, r(s_{n+1}), \phi(s_{n+1}))$ in M
    Store extra hindsight experience in M every H step
    Sample random mini-batch transitions $(\phi(s_j), a_j, r(s_{j+1}), \phi(s_{j+1}))$ of size B from M
    Set target value

$$y_j = \begin{cases} r(s_j), \text{if } s_{j+1} \text{ arrives at target;} \\ r(s_j) + \gamma Q'(\phi(s_{j+1}), \arg\max_v Q(\phi(s_{j+1}), v), \text{otherwise} \end{cases}$$

    Perform a gradient descent step on $(y_j - Q(\phi(s_j), a_j))^2$ to update the critic network parameters $\theta^Q$
    Update the actor network using the sampled policy gradient:

$$\nabla_{\theta^\mu} J \approx \frac{1}{N} \sum_i \nabla_a Q(s,a|\theta^Q)|_{s=s_i, a=\mu(s_i)} \nabla_{\theta^\mu} \mu(s|\theta^\mu)|_{s_i}$$

    Update the target networks:

$$\theta^{Q'} = \beta\theta^Q + (1-\beta)\theta^{Q'}$$
$$\theta^{\mu'} = \beta\theta^\mu + (1-\beta)\theta^{\mu'}$$

  **End For**
**End For**

---

**Table 1**. Training parameters

| Parameter | value |
| --- | --- |
| Training episode $N_E$ | ~5000 (full-control, rotor), ~20000 (translator) |
| Minibatch size, $B$ | 64 |
| Replay memory size, $N_M$ | 500000 |
| Target network update frequency $C$ | 100 |
| Discount factor, $\gamma$ | 0.99 |
| Learning rate, $\alpha$ | 0.00025 |
| Soft update parameter, $\beta$ | 0.01 |
| OU process mean level | 0 |
| OU process volatility | 0.5 |
| OU process mean reversion speed | 0.15 |
| Initial target threshold, $T_s$ | 0.1 |
| Final target threshold, $T_e$ | 1 |
| Target threshold decay, $T_d$ | ~5000 (full-control, rotor), ~20000 (translator) |
| Max step in an episode, maxStep | 100 (full-control, rotor), 500 (translator) |
| Sensor window size W | 30 |

## 5.2 Simulation setup and performance evaluation

### 5.2.1 Free space navigation and localization

For full-control and rotor motors, motors are controlled to navigate to targets with relative coordinates of (10, 0), (10, 10), (0, 10), (-10, 10), and (-10, 0), with and without Brownian motion applied. For translator motors, motors are controlled to navigate to targets with relative coordinates of (10, 0), (10, 10), (0, 10), (-10, 10), (-10, 0), (-10, 10), (0, -10), and (10, -10), with Brownian motion applied. The mean traveled distance versus time for different motors were measured from 100 navigation trajectories starting from an initial state (0, 0, 0) to a target located at (1000, 0). The localization error versus disturbance strength λ is conducted at λ = 0, 0.194, 0.274, 0.434, 0.613, 0.867, 1.171, and 1.939. The steady state simulations lasted for 3000τ to collect sufficient data samples.

*5.2.2 Navigation in obstacle environment*

In the square obstacle channel, motors are controlled to navigate from state (14, 5, 0) to a target located at (14, 105). In the cross obstacle channel, motors are controlled to navigate from an initial state (14, 5, 0) to a target located at (14, 115). The mean traveled distance versus time for different motors are measured from 100 navigation trajectories starting from an initial state (14, 5, 0) to a target located at (1000, 0). The first passage time distribution is constructed from 1000 navigation trajectories starting an initial state (14, 5, 0) towards a target located at (14, 105).

*5.2.3 Navigation with arrival time constraint*

The first passage time distribution is constructed from 1000 navigation trajectories starting from an initial state (0, 0, 0) towards a target located at (20, 20).

## Conflict of interest

The authors declare that they have no conflict of interest associated with the presented work.

## Acknowledgements

Financial support from the National Natural Science Foundation of China (11961131005, 11922207) is acknowledged.

## References

1       Mallouk, T. E. & Sen, A. Powering Nanorobots. *Scientific American* **300**, 72-77 (2009).

2       Wang, J. & Gao, W. Nano/microscale motors: biomedical opportunities and challenges. *ACS Nano* **6**, 5745-5751 (2012).

3       Li, J., Rozen, I. & Wang, J. Rocket Science at the Nanoscale. *ACS Nano* **10**, 5619-5634, doi:10.1021/acsnano.6b02518 (2016).

4       Li, J., de Ávila, B. E.-F., Gao, W., Zhang, L. & Wang, J. Micro/nanorobots for biomedicine: Delivery, surgery, sensing, and detoxification. *Science Robotics* **2**, eaam6431 (2017).

5       Sánchez, S., Soler, L. & Katuri, J. Chemically Powered Micro- and Nanomotors. *Angewandte Chemie International Edition* **54**, 1414-1444, doi:10.1002/anie.201406096 (2015).

6       Ebbens, S. J. & Gregory, D. A. Catalytic Janus Colloids: Controlling Trajectories of Chemical Microswimmers. *Acc. Chem. Res.* **51**, 1931-1939 (2018).

7       Koman, V. B. *et al.* Colloidal nanoelectronic state machines based on 2D materials for

| | |
|---|---|
| | aerosolizable electronics. *Nature Nanotechnology* **13**, 819-827 (2018). |
| 8 | Wu, Z. *et al.* A microrobotic system guided by photoacoustic computed tomography for targeted navigation in intestines in vivo. *Science Robotics* **4**, eaax0613, doi:10.1126/scirobotics.aax0613 (2019). |
| 9 | Li, J., de Ávila, B. E.-F., Gao, W., Zhang, L. & Wang, J. Micro/nanorobots for biomedicine: Delivery, surgery, sensing, and detoxification. *Science Robotics* **2**, eaam6431 (2017). |
| 10 | Soler, L., Magdanz, V., Fomin, V. M., Sanchez, S. & Schmidt, O. G. Self-propelled micromotors for cleaning polluted water. *ACS Nano* **7**, 9611-9620 (2013). |
| 11 | Goodrich, C. P. & Brenner, M. P. Using active colloids as machines to weave and braid on the micrometer scale. *Proceedings of the National Academy of Sciences* **114**, 257, doi:10.1073/pnas.1608838114 (2017). |
| 12 | Xiao, M. *et al.* pH-Responsive On-Off Motion of a Superhydrophobic Boat: Towards the Design of a Minirobot. *Small* **10**, 859-865 (2014). |
| 13 | Venugopalan, P. L. *et al.* Conformal cytocompatible ferrite coatings facilitate the realization of a nanovoyager in human blood. *Nano letters* **14**, 1968-1975 (2014). |
| 14 | Dey, K. K. *et al.* Micromotors powered by enzyme catalysis. *Nano letters* **15**, 8311-8315 (2015). |
| 15 | Medina-Sánchez, M., Schwarz, L., Meyer, A. K., Hebenstreit, F. & Schmidt, O. G. Cellular cargo delivery: Toward assisted fertilization by sperm-carrying micromotors. *Nano letters* **16**, 555-561 (2015). |
| 16 | Yang, G.-Z. *et al.* The grand challenges of Science Robotics. *Science robotics* **3**, eaar7650 (2018). |
| 17 | Guix, M., Mayorga-Martinez, C. C. & Merkoçi, A. Nano/micromotors in (bio) chemical science applications. *Chemical reviews* **114**, 6285-6322 (2014). |
| 18 | Kline, T. R., Paxton, W. F., Mallouk, T. E. & Sen, A. Catalytic nanomotors: remote-controlled autonomous movement of striped metallic nanorods. *Angewandte Chemie International Edition* **44**, 744-746 (2005). |
| 19 | Brooks, A. M., Sabrina, S. & Bishop, K. J. Shape-directed dynamics of active colloids powered by induced-charge electrophoresis. *Proceedings of the National Academy of Sciences* **115**, E1090-E1099 (2018). |
| 20 | Palagi, S. & Fischer, P. Bioinspired microrobots. *Nature Reviews Materials* **3**, 113 (2018). |
| 21 | Sánchez, S., Soler, L. & Katuri, J. Chemically powered micro-and nanomotors. *Angewandte Chemie International Edition* **54**, 1414-1444 (2015). |
| 22 | Howse, J. R. *et al.* Self-motile colloidal particles: from directed propulsion to random walk. *Physical review letters* **99**, 048102 (2007). |
| 23 | Bechinger, C. *et al.* Active particles in complex and crowded environments. *Reviews of Modern Physics* **88**, 045006 (2016). |
| 24 | Maier, A. M. *et al.* Magnetic propulsion of microswimmers with DNA-based flagellar bundles. *Nano letters* **16**, 906-910 (2016). |
| 25 | Selmke, M., Khadka, U., Bregulla, A. P., Cichos, F. & Yang, H. Theory for controlling individual self-propelled micro-swimmers by photon nudging I: directed transport. *Physical Chemistry Chemical Physics* **20**, 10502-10520, doi:10.1039/C7CP06559K (2018). |
| 26 | Selmke, M., Khadka, U., Bregulla, A. P., Cichos, F. & Yang, H. Theory for controlling individual self-propelled micro-swimmers by photon nudging II: confinement. *Physical Chemistry Chemical Physics* **20**, 10521-10532, doi:10.1039/C7CP06560D (2018). |

27  Yang, Y. & Bevan, M. A. Optimal Navigation of Self-Propelled Colloids. *ACS Nano* **12**, 10712-10724, doi:10.1021/acsnano.8b05371 (2018).

28  Yang, Y., Bevan, M. A. & Li, B. Efficient Navigation of Colloidal Robots in an Unknown Environment via Deep Reinforcement Learning. *Advanced Intelligent Systems* **0**, 1900106, doi:10.1002/aisy.201900106 (2019).

29  Silver, D. *et al.* Mastering the game of Go with deep neural networks and tree search. *Nature* **529**, 484-489 (2016).

30  Mnih, V. *et al.* Human-level control through deep reinforcement learning. *Nature* **518**, 529-533 (2015).

31  LeCun, Y., Bengio, Y. & Hinton, G. Deep learning. *Nature* **521**, 436–444, doi:10.1038/nature14539 (2015).

32  Palacci, J., Sacanna, S., Steinberg, A. P., Pine, D. J. & Chaikin, P. M. Living crystals of light-activated colloidal surfers. *Science* **339**, 936-940 (2013).

33  Sutton, R. S. & Barto, A. G. *Reinforcement Learning: An Introduction*. Vol. 1 (MIT Press, 1998).

34  Lillicrap, T. P. *et al.* Continuous control with deep reinforcement learning. *arXiv preprint arXiv:1509.02971* (2015).

35  Van Hasselt, H., Guez, A. & Silver, D. in *Thirtieth AAAI Conference on Artificial Intelligence.*

36  Andrychowicz, M. *et al.* in *Adv. Neural Inf. Process. Syst.*  5048-5058.

37  Mano, T., Delfau, J.-B., Iwasawa, J. & Sano, M. Optimal run-and-tumble–based transportation of a Janus particle with active steering. *Proceedings of the National Academy of Sciences* **114**, E2580-E2589 (2017).

38  Haeufle, D. F. B. *et al.* External control strategies for self-propelled particles: Optimizing navigational efficiency in the presence of limited resources. *Physical Review E* **94**, 012617 (2016).

39  Ron, J. E., Pinkoviezky, I., Fonio, E., Feinerman, O. & Gov, N. S. Bi-stability in cooperative transport by ants in the presence of obstacles. *PLoS computational biology* **14** (2018).

40  Feinerman, O., Pinkoviezky, I., Gelblum, A., Fonio, E. & Gov, N. S. The physics of cooperative transport in groups of ants. *Nature Physics* **14**, 683-693 (2018).

41  Yang, Y. & Bevan, M. A. Cargo capture and transport by colloidal swarms. *Science Advances* **6**, eaay7679 (2020).

42  van der Maaten, L. & Hinton, G. Visualizing data using t-SNE. *Journal of Machine Learning Research* **9**, 2579-2605 (2008).

43  Yang, Y. & Bevan, M. A. Interfacial colloidal rod dynamics: Coefficients, simulations, and analysis. *The Journal of Chemical Physics* **147**, 054902 (2017).

44  Bitter, J. L., Yang, Y., Duncan, G., Fairbrother, H. & Bevan, M. A. Interfacial and confined colloidal rod diffusion. *Langmuir* **33**, 9034-9042 (2017).

45  Liu, Y., Yang, Y., Li, B. & Feng, X.-Q. Collective oscillation in dense suspension of self-propelled chiral rods. *Soft Matter* **15**, 2999-3007 (2019).

46  Yang, Y. *Stochastic Modeling and Optimal Control for Colloidal Organization, Navigation, and Machines*, Johns Hopkins University, (2017).

47  Tang, X. *et al.* Optimal feedback controlled assembly of perfect crystals. *ACS Nano* **10**, 6791-6798 (2016).

48  Bevan, M. A. *et al.* Controlling assembly of colloidal particles into structured objects: Basic strategy and a case study. *Journal of Process Control* **27**, 64-75 (2015).

49  Edwards, T. D., Yang, Y., Beltran-Villegas, D. J. & Bevan, M. A. Colloidal crystal grain boundary formation and motion. *Scientific reports* **4**, 6132 (2014).

50  Yang, Y., Thyagarajan, R., Ford, D. M. & Bevan, M. A. Dynamic colloidal assembly pathways via low dimensional models. *The Journal of chemical physics* **144**, 204904 (2016).

51  Puterman, M. L. *Markov Decision Processes: Discrete Stochastic Dynamic Programming*. (Wiley, 2014).

52  Yang, Y. & Li, B. A simulation algorithm for Brownian dynamics on complex curved surfaces. *The Journal of chemical physics* **151**, 164901 (2019).

53  Florensa, C., Held, D., Wulfmeier, M., Zhang, M. & Abbeel, P. Reverse curriculum generation for reinforcement learning. *arXiv preprint arXiv:1707.05300* (2017).